\documentclass[aps,prb,showpacs,floatfix,twocolumn]{revtex4}
\usepackage{amsmath,amssymb,amsfonts}
\usepackage{graphics}

\begin{document}

\title{Reverse Monte Carlo Simulations, Raman Scattering and Thermal studies of an 
Amorphous Ge$_{30}$Se$_{70}$ Alloy Produced by Mechanical Alloying}

\author{K. D. Machado}
\email{kleber@fisica.ufsc.br}
\affiliation{Departamento de F\'{\i}sica, Universidade Federal de Santa Catarina, 88040-900 
Florian\'opolis, SC, Brazil}

\author{J. C. de Lima}
\affiliation{Departamento de F\'{\i}sica, Universidade Federal de Santa Catarina, 88040-900 
Florian\'opolis, SC, Brazil}

\author{C. E. M. Campos}
\affiliation{Departamento de F\'{\i}sica, Universidade Federal de Santa Catarina, 88040-900 
Florian\'opolis, SC, Brazil}

\author{T. A. Grandi}
\affiliation{Departamento de F\'{\i}sica, Universidade Federal de Santa Catarina, 88040-900 
Florian\'opolis, SC, Brazil}

\author{P. S. Pizani}
\affiliation{Departamento de F\'{\i}sica, Universidade Federal de S\~ao Carlos, 13565-905,  
S\~ao Carlos, SP, Brazil}

\date{\today}

\begin{abstract}

The short and intermediate range order of an amorphous Ge$_{30}$Se$_{70}$ alloy  
produced by Mechanical Alloying were studied by Reverse Monte Carlo simulations of its x-ray total 
structure factor, Raman scattering and differential scanning calorimetry. The simulations were 
used to compute 
the $G^{\text{RMC}}_{\text{Ge-Ge}}(r)$, $G^{\text{RMC}}_{\text{Ge-Se}}(r)$ and 
$G^{\text{RMC}}_{\text{Se-Se}}(r)$ partial distribution functions and 
the ${\cal S}^{\text{RMC}}_{\text{Ge-Ge}}(K)$, ${\cal S}^{\text{RMC}}_{\text{Ge-Se}}(K)$ and 
${\cal S}^{\text{RMC}}_{\text{Se-Se}}(K)$ partial structure factors. We calculated 
the coordination numbers and interatomic distances for 
the first and second neighbors and the bond-angle distribution functions $\Theta_{ijl}(\cos\theta)$. 
The data obtained indicate that the structure of the alloy has important differences when compared 
to alloys prepared by other techniques. There are a high number of 
Se-Se pairs in the first shell, 
and some of the tetrahedral units formed seemed to be connected by Se-Se bridges.

\end{abstract}


\maketitle

\section{Introduction}

Amorphous semiconductor materials like chalcogenide glasses present a great potential for 
application in technological devices, such as optical fibers, memory materials and switching 
devices, but their use is limited due to several factors. One of them is the difficulty in  
obtaining information about atomic structures, which define the short-range order (SRO) 
of the alloy. In this context, the 
structures of amorphous Ge$_x$Se$_{1-x}$ ({\em a}-Ge$_x$Se$_{1-x}$) and liquid 
Ge$_x$Se$_{1-x}$ ({\em l}-Ge$_x$Se$_{1-x}$), in particular Ge$_{33}$Se$_{67}$ (GeSe$_2$), 
have been extensively 
studied by several experimental techniques, like neutron diffraction (ND) 
\cite{Rao,Penfold,Salmon2,Petri}, 
x-ray diffraction (XRD) \cite{Tani}, extended x-ray absorption fine structure (EXAFS) 
\cite{Egil,Zhou} and Raman spectroscopy (RS) \cite{Hideo,Sugai}. On the theoretical side, molecular 
dynamics simulations (MD) \cite{Vashishtaprl,Vashishtaprb,Carlo,Carlo2,Cobb2,Cobb,Drabold} have been 
carried out to 
understand the SRO in these liquids and glasses in terms of two possible and distinct models. In 
the first one the distribution of 
bonds in the structure is purely randomic and determined by the local 
coordination numbers and composition. In the second one, there is a strong SRO and the structure 
is formed by well defined structural units, e.g., corner-sharing  
GeSe$_{4/2}$ (CS) tetrahedral and edge-sharing Ge$_2$Se$_{8/2}$ (ES) bitetrahedral units. The 
distribution of these units gives raise to a medium, or intermediate, range order (IRO), whose 
signature is the appearance of a first sharp diffraction peak (FSDP) in the neutron 
\cite{Petri,Rao,Salmon2} 
or x-ray structure factors \cite{Tani} at many compositions. In particular, ND experiments 
performed on melt-quenched \cite{Petri,Salmon2} (MQ) GeSe$_2$ (MQ-GeSe$_2$) showed a FSDP in the 
total structure factor ${\cal S}(K)$ which was associated with correlations in 
the range of 5--6~\AA. As described in Ref.~\onlinecite{Petri}, this alloy is formed by CS 
and ES units with heteropolar bonds but there are homopolar bonds in very small quantities. These 
results are reproduced by MD simulations in {\em l}-GeSe$_2$ \cite{Carlo,Carlo2,Cobb} 
and {\em a}-GeSe$_2$ \cite{Vashishtaprl,Vashishtaprb,Cobb2,Drabold}, except the FSDP in the 
Bathia-Thorton (BT) \cite{bathia} concentration-concentration structure factor 
${\cal S}_{\text{CC}}(K)$. It should be noted, however, 
that almost all available data about {\em a}-Ge$_x$Se$_{1-x}$ alloys were determined 
for MQ samples, and the preparation method can affect the SRO and IRO, as it was reported 
by Takeuchi {\em et al}. \cite{Hideo} by comparing the structures of {\em a}-Ge$_{30}$Se$_{70}$ 
produced by MQ and by vacuum evaporation (VE) techniques. Tani {\em et al}. \cite{Tani}  
studied the {\em a}-GeSe$_2$ produced by Mechanical Grinding (MG) of its crystalline counterpart 
and also found some structural differences. These differences are important because some 
physicochemical properties can be altered and improved as desired by choosing an appropriate 
method of preparation. 

It is well known that the Mechanical Alloying technique (MA) \cite{Mec} introduces 
a high quantity of disorder and defects in the 
structure of the materials produced using this process. Thus, in this paper, we investigated 
the SRO and IRO of an amorphous Ge$_{30}$Se$_{70}$ alloy produced by MA (MA-{\em a}-Ge$_{30}$Se$_{70}$) 
starting from 
the elemental powders of Ge and Se using Raman spectroscopy, Differential Scanning Calorimetry 
(DSC), X-ray diffraction and reverse Monte Carlo 
(RMC) simulations \cite{RMC1,RMC2,RMCA,rmcreview} of its XRD ${\cal S}(K)$. 
We were interested in studying two main points. Since it is not obvious that starting from the 
powders of Ge and Se and submitting them to a milling process we would obtain an alloy formed by 
(ordered) structural units like CS or ES units, first of all we would like to know if the alloy 
produced by MA contains these units. Besides that, even if these units are formed the SRO and 
the IRO of the 
alloy can be significantly altered by the high quantity of defects and disorder introduced by the 
MA process when compared to MQ samples, for instance. Therefore, the second point is to determine the 
local structure of the alloy itself, finding coordination numbers and interatomic distances.
At our knowledge, this is the first 
time that such study is reported concerning an {\em a}-Ge$_{30}$Se$_{70}$ alloy produced by MA. 

\section{Theoretical Background}

\subsection{Structure Factors}

\subsubsection{Faber and Ziman structure factors}

According to Faber and Ziman \cite{Faber}, 
the total structure factor ${\cal S}(K)$ is obtained 
from the scattered intensity per atom $I_a(K)$ through

\begin{eqnarray}
{\cal S}(K) &=& \frac{I_a(K)-\bigl[\langle f^2(K)\rangle - \langle f(K)\rangle^2
\bigr]}{\langle f(K)\rangle^2} \nonumber\,,\\
&=& \sum_{i=1}^n{\sum_{j=1}^n{w_{ij}(K) {\cal S}_{ij}(K) }}\,,\nonumber
\label{eqstructurefactor}
\end{eqnarray}

\noindent where $K$ is the transferred momentum,  
${\cal S}_{ij}(K)$ are the partial structure factors and $w_{ij}(K)$ are given by

\begin{equation}
w_{ij}(K) = \frac{c_i c_j f_i(K) f_j(K)}{\langle f(K)\rangle^2}\,,\nonumber
\label{eqw}
\end{equation}
 
\noindent and

\begin{eqnarray}
\langle f^2(K) \rangle &=& \sum_{i}{ c_i f_i^2(K)}\nonumber\,,\\
\langle f(K) \rangle^2 &=& \Bigl[\sum_{i}{ c_i f_i(K)}\Bigr]^2 \nonumber\,.
\end{eqnarray}

\noindent Here, $f_i(K)= f_{0_i}(K)+f_i'(\lambda)+if_i''(\lambda)$ is the atomic scattering factor, 
$f_i'(\lambda)$ and $f_i''(\lambda)$ are the anomalous dispersion terms and $c_i$ is the concentration 
of atoms of type $i$ ($\lambda$ is the radiation wavelength). The partial reduced 
distribution functions $G_{ij}(r)$ 
are related to ${\cal S}_{ij}(K)$ through

\begin{equation}
G_{ij}(r) = \frac{2}{\pi} \int_0^{\infty}{K\bigl[{\cal S}_{ij}(K)-1 \bigr] 
\sin (Kr)\, dK}\,.\nonumber
\label{eqpartialg}
\end{equation}

\noindent From the $G_{ij}(r)$ functions the partial radial distribution function $\text{RDF}_{ij}(r)$ 
can be calculated by

\begin{equation}
\text{RDF}_{ij}(r) = 4\pi \rho_0 c_j r^2+ r G_{ij}(r)\,.\nonumber
\label{prdf}
\end{equation}

\noindent where $\rho_0$ is the density of the alloy (in atoms/\AA$^3$). Interatomic distances 
are obtained from the maxima of $G_{ij}(r)$ and 
coordination numbers are calculated by integrating the peaks of $\text{RDF}_{ij}(r)$.

\subsubsection{Bathia and Thornton structure factors}

The BT structure factors can be related to the FZ ones \cite{bathia}. 
For a binary alloy the BT number-number structure factor ${\cal S}_{\text{NN}}(K)$ is given by

\begin{equation}
{\cal S}_{\text{NN}}(K) = c_1^2 {\cal S}_{11}(K) + c_2^2 {\cal S}_{22}(K) + 
2c_1 c_2 {\cal S}_{12}(K) \,,
\label{btnn}
\end{equation}

\noindent where ${\cal S}_{ij}(K)$ are the FZ partial structure factors and $c_i$ is  
the concentration of element $i$. The BT number-concentration structure factor 
${\cal S}_{\text{NC}}(K)$ is

\begin{widetext}

\begin{equation}
{\cal S}_{\text{NC}}(K) = c_1c_2\Bigl\{c_1\bigl[{\cal S}_{11}(K) - {\cal S}_{12}(K) \bigr]
- c_2\bigl[{\cal S}_{22}(K)-{\cal S}_{12}(K)\bigr]\Bigr\}\,,
\label{btnc}
\end{equation}

\end{widetext}

\noindent and the BT concentration-concentration structure factor ${\cal S}_{\text{CC}}(K)$ 
is found through

\begin{widetext}
\begin{equation}
{\cal S}_{\text{CC}}(K) =
c_1c_2\Bigl\{1+ c_1c_2\bigl[{\cal S}_{11}(K) + {\cal S}_{22}(K) 
- 2{\cal S}_{12}(K)\bigr]\Bigr\}\,.
\label{btcc}
\end{equation}
\end{widetext}

\subsection{RMC Method}

The basic idea and the algorithm of the standard RMC method are described 
elsewhere\cite{RMC1,RMC2,RMCA,rmcreview} and its application to different materials is reported in 
the literature 
\cite{Rosi,mellergard,RMC5,RMC6,RMC7,RMC9,RMC11,RMC12,RMC13,kleber,kleber2,rmcNDXRD,Iparraguirre,Ohkubo,Svab}. 
In the RMC procedure, a three-dimensional arrangement of atoms with the same density and chemical 
composition of the alloy is placed into a cell (usually cubic) with periodic boundary conditions and 
the $G_{ij}^{\text{RMC}}(r)$ functions corresponding to it are directly calculated through

\begin{equation}
G^{\text{RMC}}_{ij}(r) = \frac{n^{\text{RMC}}_{ij}(r)}{4\pi\rho_0 r^2\Delta r}\,, \nonumber
\end{equation}

\noindent where $n^{\text{RMC}}_{ij}(r)$ is the number of atoms at a distance between $r$ and 
$r + \Delta r$ from the central atom, averaged over all atoms. By allowing the atoms to 
move (one at each time) inside the cell, the $G^{\text{RMC}}_{ij}(r)$ functions can be changed 
and, as a consequence, ${\cal S}_{ij}^{\text{RMC}}(K)$ 
and ${\cal S}^{\text{RMC}}(K)$ are changed. 
Thus, ${\cal S}^{\text{RMC}}(K)$ is compared to the ${\cal S}(K)$ factor in 
order to minimize the differences between them. The function to be minimized is 

\begin{equation}
\psi^2 = \frac{1}{\delta}\sum_{i=1}^m{\bigl[{\cal S}(K_i)-{\cal S}^{\text{RMC}}(K_i)\bigr]^2}\,,
\label{eqpsi}
\end{equation}

\noindent where the sum is over $m$ experimental points and $\delta$ is related to the 
experimental error in ${\cal S}(K)$. If the movement decreases $\psi^2$, it is always accepted. 
If it increases $\psi^2$, it is accepted with a probability given by $\exp(-\Delta 
\psi^2/2)$; otherwise it is rejected. As this process is iterated $\psi^2$ decreases until it 
reaches an equilibrium value. Thus, the atomic configuration corresponding to  
equilibrium should be consistent with the experimental total structure factor within the 
experimental error. By using the $G^{\text{RMC}}_{ij}(r)$ functions 
the coordination numbers and 
interatomic distances can be calculated. In addition, the bond-angle distributions $\Theta_{ijl}(\cos 
\theta)$ can also be determined.

\section{Experimental Procedure}

The MA-{\em a}-Ge$_{30}$Se$_{70}$ alloy was produced by considering a binary mixture of high-purity 
elemental powders of germanium (Alfa Aesar 99.999\% purity, particle size $<$ 150 $\mu$m) and 
selenium (Alfa Aesar 99.999\% purity, particle size $<$ 150 $\mu$m) that was sealed together 
with several steel balls into a 
cylindrical steel vial under an argon atmosphere. The ball-to-powder weight ratio was 5:1. A 
Spex Mixer/Mill model 8000 was used to perform MA at room temperature. 
The mixture was continuously milled for 40 h. A ventilation system was used to keep the vial 
temperature close to room temperature. The XRD pattern was recorded in 
a powder Siemens diffractometer equipped with a graphite monochromator, using 
the CuK$_\alpha$ line ($\lambda = 1.5418$ \AA). The total structure factor ${\cal S}(K)$ was 
computed from the XRD pattern after corrections for polarization, absorption, 
and inelastic scattering, following the procedure described by Wagner \cite{Wagner}. The 
$f'$ and $f''$ values were taken from a table compiled by Sasaki \cite{Sasaki}. 
Raman measurements were performed with a T64000 Jobin-Yvon triple monochromator coupled to 
a cooled CCD detector and a conventional photon counting system. The 
5145~\AA\ line of an argon ion laser was used as exciting light, always in backscattering geometry. 
The output power of the laser was kept at about 200 mW to avoid overheating the samples. All Raman 
measurements were performed at room temperature. 
DSC measurements were taken with a heating rate of 10$^\circ$C/min using a TA 2010 DSC cell 
in a flowing argon atmosphere.

\section{Results and Discussion}

\subsection{Raman Scattering}

\begin{figure}[h]
\includegraphics{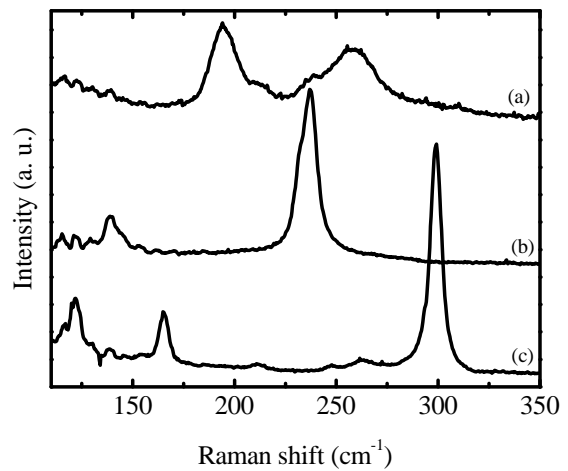}
\caption{\label{figr} RS spectra of (a) MA-{\em a}-Ge$_{30}$Se$_{70}$, 
(b) {\em c}-Se and (c) {\em c}-Ge.}
\end{figure}

Figure~\ref{figr} shows the RS spectra of MA-{\em a}-Ge$_{30}$Se$_{70}$, 
{\em c}-Se and {\em c}-Ge. The alloy has bands at around 
195, 215, 237 and 255~cm$^{-1}$. The band at 195~cm$^{-1}$ is associated with the $A_1$ 
breathing mode of CS units and the 215~cm$^{-1}$ band is related to the $A_1^c$ breathinglike 
motions (companion peak) of Se in the ES units \cite{Sugai,Yong,Popovic}. The difference in the 
intensities of these peaks indicates that the alloy is formed basically by CS tetrahedra and 
ES tetrahedra are found in a small quantity. 
The shoulder around 237~cm$^{-1}$ is associated with $A_1$ and $E$ modes of Se chains, and 
they are also seen in the {\em c}-Se (see Fig.~\ref{figr}). The broad band at 255~cm$^{-1}$ is 
related to $A_1$ and $E_2$ modes of Se$_n$ rings \cite{Yong,Popovic,Hideo,Egil,Kohara}. It is 
important to note that the band at around 165~cm$^{-1}$, which is associated with Ge-Ge pairs 
vibrations in ethanlike units, is not seen in the  
spectrum of the alloy (as well as the other bands of {\em c}-Ge), indicating that the quantity 
of Ge-Ge pairs is very low in the alloy. In addition, the bands related to Se-Se pairs are not seen in 
the spectra of the alloys produced by MQ \cite{Yong} or VE \cite{Hideo} techniques at this composition, 
but they exist in 
the alloy produced by MA, and their intensities suggest that their quantities may be relevant. 
These results 
indicate that the tetrahedral units are formed during the MA process and there is a preference for 
CS units. 

\subsection{DSC measurement}

\begin{figure}[h]
\includegraphics{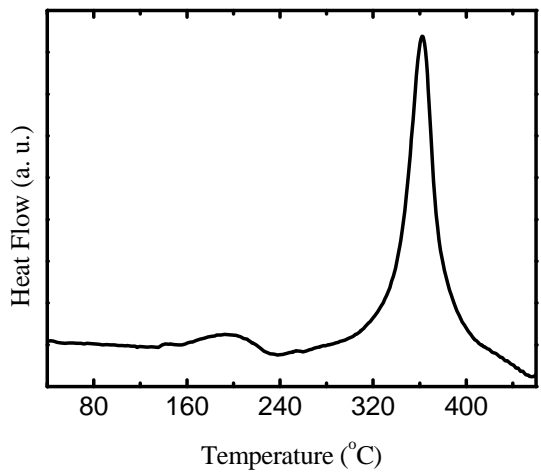}
\caption{\label{figdsc} DSC measurement performed on MA-{\em a}-Ge$_{30}$Se$_{70}$.}
\end{figure}

Figure~\ref{figdsc} shows the DSC measurement of 
the alloy, and it displays only two exothermic peaks at about 200$^\circ$C and 360$^\circ$C, which 
are associated with crystallization temperatures of 
MA-{\em a}-Ge$_{30}$Se$_{70}$. There are no peaks associated with melting of {\em c}-Se, which should 
appear around 217$^\circ$C, nor with glass transition or crystallization of {\em a}-Se, which should 
occur at about 45$^\circ$C and 90$^\circ$C, respectively \cite{Tani,Joao}. These data indicate that 
the Raman lines 
associated with Se chains and rings are not due to the presence of some quantity of 
unreacted {\em a}-Se or {\em c}-Se  
and reinforce the existence of a high number of Se-Se pairs in the alloy, indicating that 
the local structure of MA-{\em a}-Ge$_{30}$Se$_{70}$  
probably is different from that found in MQ, VE and even in MG samples, since now there are 
Se-Se pairs in a higher quantity.

\subsection{X-ray Diffraction and RMC Simulations}

\begin{figure}[h]
\includegraphics{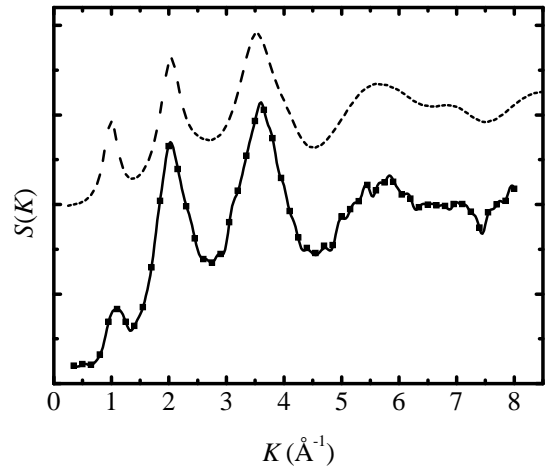}
\caption{\label{fig1} Experimental (full line) and simulated 
(squares) total structure factor for MA-{\em a}-Ge$_{30}$Se$_{70}$ together with the 
ND total structure factor given in Ref. \onlinecite{Salmon2} (dashed line) (for a better 
comparison it was cut at $K_{\text{max}} = 8.5$ \AA$^{-1}$).}
\end{figure}

Figure~\ref{fig1} shows the experimental XRD ${\cal S}(K)$ (full line) for 
our alloy and the experimental ND ${\cal S}(K)$ (dashed line) given in Refs.  \onlinecite{Petri} 
or \onlinecite{Salmon2} for {\em a}-GeSe$_2$. The FSDP is 
seen at around 1.1~\AA$^{-1}$. It is lower than those shown in Refs.~\onlinecite{Salmon2} and 
\onlinecite{Rao}, indicating that the 
IRO in the alloy produced by MA is less pronounced than in the MQ-GeSe$_2$ 
samples \cite{Petri,Rao,Salmon2}. The FSDP is known to be much dependent on Ge-Ge and, to a lesser 
extent, on Ge-Se correlations \cite{fuoss2,Moss,Vashishtaprb,Salmon2}. Therefore these correlations 
have a different behavior in MA-{\em a}-Ge$_{30}$Se$_{70}$. 

${\cal S}(K)$ was simulated using the RMC program 
available on the Internet \cite{RMCA}. To perform the simulations we have considered a cubic cell 
with 16000 atoms (4800 Ge and 11200 Se), $\delta = 0.002$, and a mean atomic number density 
$\rho_0 = 0.03868\pm 0.0005$ atoms/\AA$^3$. This value was found from the slope of the straight 
line ($-4\pi\rho_0 r$) fitting the initial part (until the first minimum) of the total 
$G(r)$ function \cite{Waseda}. 
The minimum distances between atoms were fixed at the beginning of 
the simulations at $r_{\text{Ge-Ge}} = 1.8$~\AA, $r_{\text{Ge-Se}} = 1.8$~\AA\  and $r_{\text{Se-Se}} 
= 1.75$~\AA. The ${\cal S}^{\text{RMC}}(K)$ obtained from the simulations (squares) is 
also shown in Fig. \ref{fig1} and there is a very good agreement with the experimental ${\cal S}(K)$.

First hard sphere simulations without experimental data were carried out to avoid possible memory 
effects of the initial configurations in the results. Then unconstrained runs (i.e. when only hard 
sphere diameters and experimental data were used during the simulation) were carried out. These runs 
led to essentially identical $G_{ij}^{\text{RMC}}(r)$ functions, because of the proximity of the 
atomic numbers and scattering factors of Ge and Se. In the next series of simulations we tried to 
fit the experimental ${\cal S}(K)$ using the ND coordination numbers reported in 
Refs.~\onlinecite{Petri} or \onlinecite{Salmon2} 
as coordination constraints. In this case we reached a poor agreement between 
the experimental and the simulated structure factor, indicating that the local structure of  
MA-{\em a}-Ge$_{30}$Se$_{70}$ is really different from that found in the MQ-GeSe$_2$ alloy, in 
agreement with Raman results. Thus, based on Raman results (small number of Ge-Ge pairs, presence of 
CS and ES-like units and a high number of Se-Se pairs) we used as starting coordination constraints 
the values $N_{\text{Ge-Ge}}=0.2$, $N_{\text{Ge-Se}}=3.7$ and $N_{\text{Se-Se}}=1.0$, which 
were then allowed to vary freely. The 
results obtained from the best simulation achieved are shown in Fig.~\ref{fig1} and they 
are discussed below. 
As a final test, we also tried to make simulations forcing the Se-Se coordination to be two, as it 
is in {\em a}-Se, but again the simulations did not show a good convergence. When this constraint 
is released, the Se-Se coordination number decreases until it reaches the number we found for 
the previous case ($N_{\text{Se-Se}}=1.25$, see text below), in agreement with Raman and DSC results.

\begin{figure}[b]
\includegraphics{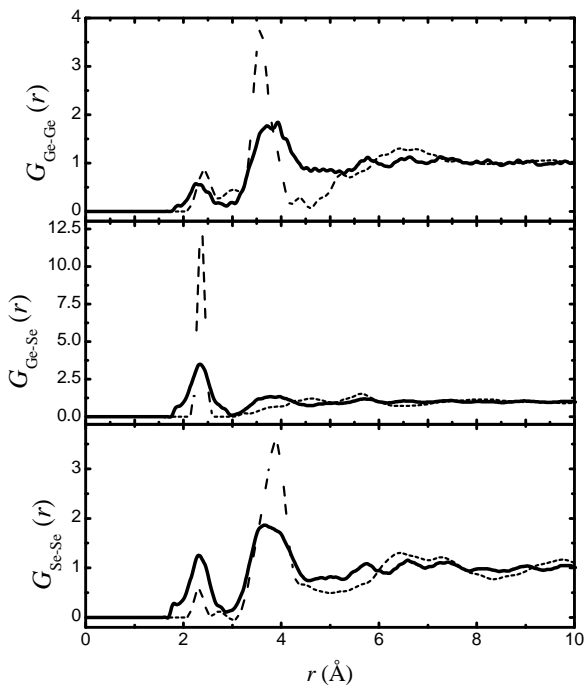}
\caption{\label{fig2} $G_{\text{Ge-Ge}}(r)$, 
$G_{\text{Ge-Se}}(r)$ and $G_{\text{Se-Se}}(r)$ 
functions obtained from RMC simulations (solid lines) and by ND (dashed lines, 
Ref. \onlinecite{Salmon2}).}
\end{figure}

\subsubsection{Pair Distribution Functions}

Figure~\ref{fig2} shows the $G_{ij}^{\text{RMC}}(r)$ functions obtained from the RMC 
simulations of MA-{\em a}-Ge$_{30}$Se$_{70}$ compared to those found for MQ-GeSe$_2$ 
\cite{Salmon2}. There are important differences among these 
functions and those obtained by ND measurements \cite{Petri,Salmon2} and also by MD simulations 
\cite{Vashishtaprl,Vashishtaprb,Cobb2,Drabold}. First of all, we should note that the densities of 
the alloys are very different ($\rho_{\text{MQ-GeSe$_2$}} = 0.0334$ atoms/\AA$^{-3}$\ ~\cite{Azoulay}), 
so the heights of the peaks in $G_{ij}(r)$ cannot be 
easily compared. The intensity of the first peak in 
$G_{\text{Se-Se}}^{\text{RMC}}$ is higher, confirming their existence in a larger quantity in the 
first coordination shell, as indicated by the RS measurement. The two first peaks of the 
$G_{\text{Ge-Ge}}^{\text{RMC}}(r)$ 
function, which correspond to Ge-Ge first and second neighbors, show up around 2.33 and 3.83~\AA. 
The first peak occurs at a distance a little shorter than that found in the MQ sample 
\cite{Petri,Salmon2}, 
but the second is displaced towards higher-{\em r} values by 0.26~\AA. Besides that, a 
minimum occurs at 
around 3.0~\AA. Remembering that the distance between two Ge atoms in adjacent ES and CS units are 
found at 3.02~\AA\ and 3.60~\AA, respectively \cite{Cobb2,Vashishtaprb}, it can be seen that the 
fraction of ES units in 
MA-{\em a}-Ge$_{30}$Se$_{70}$ is low, in agreement with the results obtained by RS. 
If we 
consider that in {\em c}-GeSe$_2$ each ES or CS Ge atom has, respectively, three and four 
nearest-neighbor Ge atoms and there are equal numbers of ES and CS units giving $N_{\text{Ge-Ge}} 
= 3.5$ we find that, for a value of $N_{\text{Ge-Ge}} = 3.85$ (see Table~\ref{tabI}), we should 
have about 
85\% of the tetrahedral units as CS units and 15\% as ES units. Since the intensity of the FSDP 
in ${\cal S}(K)$ seems to be related to the quantity of ES tetrahedra 
\cite{Vashishtaprl,Vashishtaprb,Cobb2,Cobb}, the low quantity of ES units in our alloy could 
explain the low intensity of the FSDP.

The first peak of $G_{\text{Ge-Se}}^{\text{RMC}}(r)$ function is located at 2.35~\AA, as it 
happens to the MQ-GeSe$_2$ samples \cite{Petri,Salmon2,Vashishtaprb}. This shell in our alloy 
is lower and broader than in MQ-GeSe$_2$ but, due to the difference in densities, Ge-Se coordination 
numbers are almost the same in both alloys. 
The next peak appears at 3.84~\AA, and 
it is higher than that at 4.96~\AA. In the MQ-GeSe$_2$ samples, there is a peak around 3.02~\AA\ 
which is not seen, or at least not resolved, in MA-{\em a}-Ge$_{30}$Se$_{70}$, and there  
are peaks at 3.78~\AA\ (smaller) and 4.66~\AA\ (higher). In {\em c}-GeSe$_2$ an ES Ge atom 
has Se neighbors in the range $4.6 \alt r \alt 5.3$~\AA, and a CS Ge atom has Se neighbors in 
the range $4.0 \alt r \alt 4.8$~\AA. \cite{Penfold,Rao} Then, we have 
associated the peak at 3.84~\AA\ 
with CS units and that at 4.96~\AA\ with CS and ES units.

The first peak of $G_{\text{Se-Se}}^{\text{RMC}}(r)$ function is located at 2.33~\AA, and 
it corresponds to a coordination number $N_{\text{Se-Se}} = 1.25$, which is much higher 
than that obtained for the MQ-GeSe$_2$ samples ($N_{\text{Se-Se}} = 0.20$) \cite{Petri,Salmon2}. 
This suggests that some of the tetrahedral  
units are connected by Se ``bridges", forming sequences such as Ge-Se-Se-Se-Ge. As a 
consequence, Ge-Ge pairs should be found at higher distance values. This agrees with the 
results obtained from the $G_{\text{Ge-Ge}}^{\text{RMC}}(r)$ function. The next peak 
appears at 3.75~\AA, which gives a ratio of Ge-Se:Se-Se distances of 0.627. The 
value expected for ideal tetrahedral coordination is $\sqrt{3/8} = 0.612$, indicating that 
the tetrahedral (CS or ES) units are distorted in our alloy. 
It is interesting to note that in MQ-GeSe$_2$ samples 
no peaks are found from $\approx 4.5$~\AA\ to $\approx 5.9$~\AA\ either considering 
ND results \cite{Petri,Salmon2} or MD simulations \cite{Vashishtaprb,Drabold}. On the other hand, in  
MA-{\em a}-Ge$_{30}$Se$_{70}$ there are peaks at 4.95 and 5.75~\AA, and we believe these peaks 
are related to the distances between Se atoms in the ``bridges" and Se atoms in tetrahedral units. 
The peaks at around 6.6 and 
7.4~\AA, which are also seen in MQ-GeSe$_2$ samples, can be associated, following 
Ref.~\onlinecite{Vashishtaprb}, to distances between 
Se atoms belonging to two adjacent tetrahedral units.

\begin{figure}[h]
\includegraphics{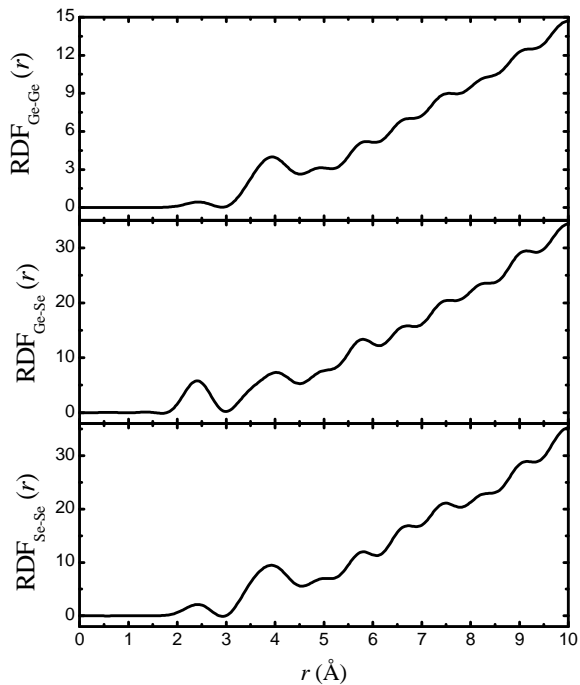}
\caption{\label{fig3} RDF$^{\text{RMC}}_{\text{Ge-Ge}}(r)$, RDF$^{\text{RMC}}_{\text{Ge-Se}}(r)$ 
and RDF$^{\text{RMC}}_{\text{Se-Se}}(r)$ obtained from the RMC simulations.}
\end{figure}

The coordination numbers show in Table~\ref{tabI} were calculated considering the 
RDF$_{ij}^{\text{RMC}}(r)$ functions shown in Fig.~\ref{fig3}. The integrations were made using the 
following ranges: from 1.7~\AA\ to 2.95~\AA\ to the first peak and from 2.95~\AA\ to 4.5~\AA\ 
to the second peak, for all RDF$_{ij}^{\text{RMC}}(r)$ functions. The interatomic distances are 
also shown in Table~\ref{tabI}.

\begingroup
\squeezetable
\begin{table}
\caption{\label{tabI} Structural Parameters obtained for MA-{\em a}-Ge$_{30}$Se$_{70}$.}
\begin{ruledtabular}
\begin{tabular}{ccccc|cccc}
\multicolumn{9}{c}{}\\[-0.4cm]
\multicolumn{9}{c}{RMC} \\[0.1cm]\hline\hline
& \multicolumn{2}{c}{} & \multicolumn{2}{c|}{} & \multicolumn{2}{c}{} & 
\multicolumn{2}{c}{}\\[-0.3cm]
& \multicolumn{4}{c|}{First Shell} & \multicolumn{4}{c}{Second Shell}\\[0.1cm]\hline
& \multicolumn{2}{c}{} & \multicolumn{2}{c|}{} & \multicolumn{2}{c}{} & 
\multicolumn{2}{c}{}\\[-0.3cm]
Bond Type & Ge-Ge & Ge-Se & Se-Ge & Se-Se & Ge-Ge & Ge-Se & Se-Ge & Se-Se \\[0.1cm]
$N$ & 0.26 & 3.50 & 1.75 & 1.25 & 3.85 & 7.4 & 3.7 & 9.7 \\
$r$ (\AA) & 2.33 & 2.35 & 2.35 & 2.33 & 3.83 & 3.84 & 3.84 & 3.75 \\\hline\hline
\multicolumn{9}{c}{}\\[-0.3cm]
\multicolumn{9}{c}{MQ-GeSe$_2$ studied by ND \cite{Petri,Salmon2}} \\[0.1cm]\hline\hline
& \multicolumn{2}{c}{} & \multicolumn{2}{c|}{} & \multicolumn{2}{c}{} & 
\multicolumn{2}{c}{}\\[-0.3cm]
Bond Type & Ge-Ge & Ge-Se & Se-Ge & Se-Se  & Ge-Ge & Ge-Se\footnote{These numbers are not 
given in Refs.~\onlinecite{Petri} or \onlinecite{Salmon2}.} & 
Se-Ge$^{\text{{\em a}}}$ & Se-Se \\[0.1cm]
$N$ & 0.25 & 3.7 & 1.8 & 0.20 & 3.2 & - & - & 9.3 \\
$r$ (\AA) & 2.42 & 2.36 & 2.36 & 2.32 & 3.57 & - & - & 3.89 \\\hline\hline
\multicolumn{9}{c}{}\\[-0.3cm]
\multicolumn{9}{c}{{\em l}-GeSe$_2$ studied by ND \cite{Penfold}} \\[0.1cm]\hline\hline
& \multicolumn{2}{c}{} & \multicolumn{2}{c|}{} & \multicolumn{2}{c}{} & 
\multicolumn{2}{c}{}\\[-0.3cm]
Bond Type & Ge-Ge & Ge-Se & Se-Ge & Se-Se  & Ge-Ge & Ge-Se & Se-Ge & Se-Se \\[0.1cm]
$N$ & 0.25 & 3.5 & 1.7 & 0.23 & 2.9 & 4.0 & 2.0 & 9.6 \\
$r$ (\AA) & 2.33 & 2.42 & 2.42 & 2.30 & 3.59 & 4.15 & 4.15 & 3.75 \\[-0.06cm]
%& \multicolumn{2}{c}{} & \multicolumn{2}{c|}{} & \multicolumn{2}{c}{} & 
%\multicolumn{2}{c}{}\\[-0.5cm]
\end{tabular}
\end{ruledtabular}
\end{table}
\endgroup

\subsubsection{Partial Structure Factors}

\begin{figure}[h]
\includegraphics{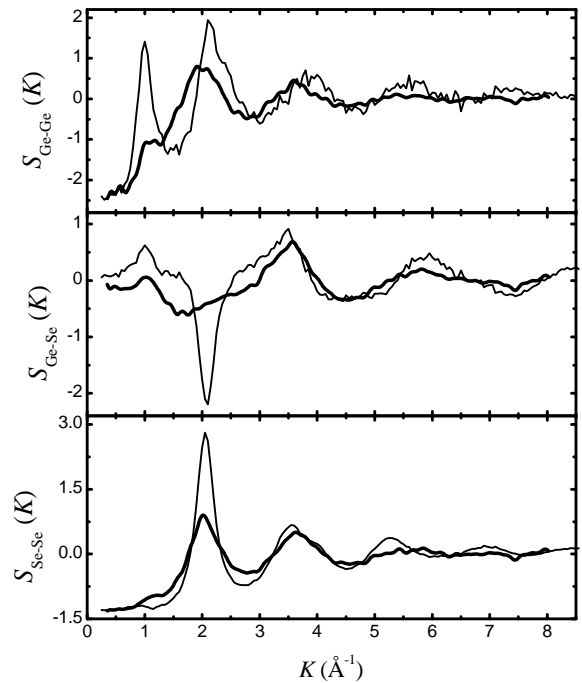}
\caption{\label{fig4} ${\cal S}_{\text{Ge-Ge}}(K)$, 
${\cal S}_{\text{Ge-Se}}(K)$ and ${\cal S}_{\text{Se-Se}}(K)$ 
factors obtained from the RMC simulations (thick lines) and by ND (thin lines, Ref. 
\onlinecite{Salmon2}).}
\end{figure}

The partial ${\cal S}_{ij}^{\text{RMC}}(K)$ are shown in Fig.~\ref{fig4}, together with 
the ${\cal S}_{ij}^{\text{ND}}(K)$ found in Ref. \onlinecite{Salmon2}. 
${\cal S}_{\text{Ge-Ge}}^{\text{RMC}}(K)$ has its first three peaks at about 1.1, 2.0 and 
3.6~\AA$^{-1}$ and two minima at 1.3 and 2.8~\AA$^{-1}$. Their positions agree reasonably 
well with those found for MQ-GeSe$_2$ studied by ND \cite{Petri,Salmon2}, but intensities 
are very different. The first peak is much lower in MA-{\em a}-Ge$_{30}$Se$_{70}$ than 
it is in the MQ-GeSe$_2$ sample. This peak is associated with the FSDP in the ${\cal S}(K)$ 
shown in Fig.~\ref{fig1} and, since the FSDP is known to be strongly dependent on 
the Ge-Ge correlations and, to a lesser extent, on the Ge-Se correlations 
\cite{fuoss2,Moss,Vashishtaprb}, the low intensity FSDP experimentally observed in Fig.~\ref{fig1} 
could be caused by weak Ge-Ge and Ge-Se correlations at its position. 
In addition, the heights of the second and third peaks are almost the same, and in MQ samples 
the height of the second peak is twice of that of the third peak. It is interesting to note 
that the first peak obtained from the MD simulations \cite{Vashishtaprl,Vashishtaprb} was found 
only at 1.3~\AA$^{-1}$, where our ${\cal S}_{\text{Ge-Ge}}^{\text{RMC}}(K)$ has a minimum.

${\cal S}_{\text{Ge-Se}}^{\text{RMC}}(K)$ has two maxima at 1.02 and 3.58~\AA$^{-1}$ 
and a minimum at 1.7~\AA$^{-1}$. Again, the first peak of MA-{\em a}-Ge$_{30}$Se$_{70}$ 
is lower than that found in MQ-GeSe$_2$ samples, but its position is the same, and it is also associated 
with the FSDP in ${\cal S}(K)$. The second peak and 
the first minimum in MQ samples are found at 3.5~\AA$^{-1}$ and 2.1~\AA$^{-1}$, respectively, 
indicating that in our alloy they are dislocated to higher (the maxima) and lower-$K$ (the minimum) 
values. These facts can be explained by 
the important differences between the $G_{\text{Ge-Se}}^{\text{RMC}}(r)$ function and that of the MQ 
samples for $r > 4$~\AA.

${\cal S}_{\text{Se-Se}}^{\text{RMC}}(K)$ has two peaks at around 2.0 
and 3.6~\AA$^{-1}$, and there is a minimum at 2.8~\AA$^{-1}$. At about 1.2~\AA$^{-1}$ there is a 
small peak associated with the FSDP in ${\cal S}(K)$. In MQ-GeSe$_2$ samples 
the peaks are seen at 0.95, 2.05 and 3.55~\AA$^{-1}$, and there is a minimum at 2.75~\AA$^{-1}$. These 
data indicate that ${\cal S}_{\text{Se-Se}}^{\text{RMC}}(K)$ is similar to 
that of MQ-GeSe$_2$ samples, at least in the low-$K$ region, concerning peak positions. However, 
we should note that their heights are different, in particular the intensity of the peak at 
2.0 \AA$^{-1}$.

\begin{figure}[h]
\includegraphics{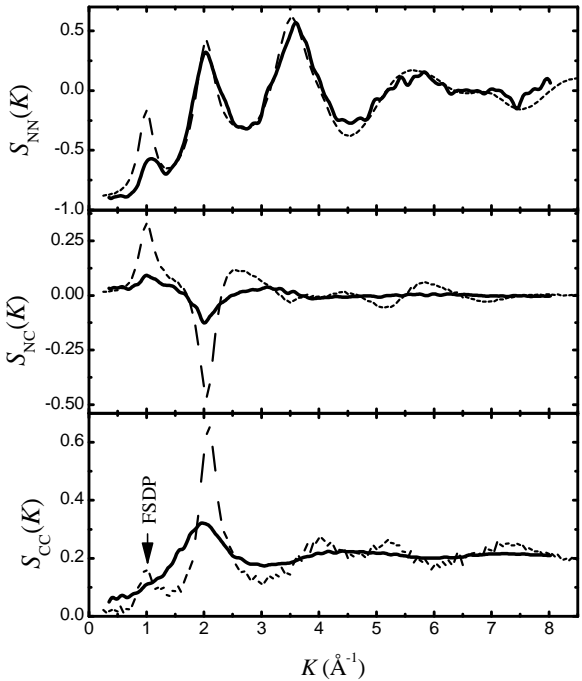}
\caption{\label{fig6} Bathia-Thornton ${\cal S}_{\text{NN}}(K)$, 
${\cal S}_{\text{NC}}(K)$ and ${\cal S}_{\text{CC}}(K)$ factors obtained from RMC 
simulations (solid lines) and by ND (dashed lines, Ref. \onlinecite{Salmon2}). The arrow indicates the 
FSDP in ${\cal S}_{\text{CC}}^{\text{RMC}}(K)$.}
\end{figure}

It is important to compare our results considering the BT formalism. Figure~\ref{fig6} 
shows the ${\cal S}_{\text{NN}}^{\text{RMC}}(K)$, ${\cal S}_{\text{NC}}^{\text{RMC}}(K)$ 
and ${\cal S}_{\text{CC}}^{\text{RMC}}(K)$ 
factors obtained using Eqs.~\ref{btnn}, \ref{btnc} and~\ref{btcc}, together with those 
factors found by ND and shown in Ref. \onlinecite{Salmon2}. As expected, 
${\cal S}_{\text{NN}}^{\text{RMC}}(K)$ resembles the XRD ${\cal S}(K)$ because the scattering 
lengths of 
Ge and Se are almost the same, and it is very similar to the ${\cal S}_{\text{NN}}^{\text{ND}}(K)$ 
found for MQ-GeSe$_2$, except for the FSDP intensity at 1.0 \AA$^{-1}$. 
Although there are differences in the peak intensities, 
${\cal S}_{\text{NC}}^{\text{RMC}}(K)$ resembles that found 
for {\em l}-GeSe$_2$ \cite{Penfold,Carlo,Carlo2,Salmon} and for 
MQ-GeSe$_2$\ ~\cite{Salmon2}, including the sharp minimum at 
2.0~\AA$^{-1}$. 
${\cal S}_{\text{CC}}(K)$ also behaves like that obtained for 
{\em l}-GeSe$_2$\ ~\cite{Penfold,Salmon} and MQ-GeSe$_2$\ ~\cite{Salmon2}. In 
{\em l}-GeSe$_2$ a sharp FSDP is clearly seen in ${\cal S}_{\text{CC}}(K)$ at around 
1.0~\AA$^{-1}$, and this fact 
also occurs in MQ-GeSe$_2$ samples. MD simulations of 
{\em l}-GeSe$_2$\ ~\cite{Carlo,Carlo2} and MQ-GeSe$_2$\ ~\cite{Drabold} could not reproduce 
well this peak. In our case, a very weak FSDP can be seen at 
1.1~\AA$^{-1}$ as a shoulder of the high peak at 2.0~\AA$^{-1}$. Remembering that the FSDP in 
${\cal S}(K)$ (see Fig.~\ref{fig1}) has a low intensity and considering all the 
results discussed above, we should not expect that the FSDP in ${\cal S}_{\text{CC}}(K)$ 
was as high and well defined as it is in {\em l}- or MQ-GeSe$_2$ samples. The 
IRO in MA-{\em a}-Ge$_{30}$Se$_{70}$ is 
different mainly because of the introduction of Se-Se first neighbor pairs as ``bridges" between 
the tetrahedral units, thus decreasing the number of ES units and increasing the number of 
CS units. This affects the short and medium 
range order, which, in its turn, 
changes the concentration-concentration BT factor.

\subsubsection{Bond-Angle Distribution Functions}

\begin{figure}[h]
\includegraphics{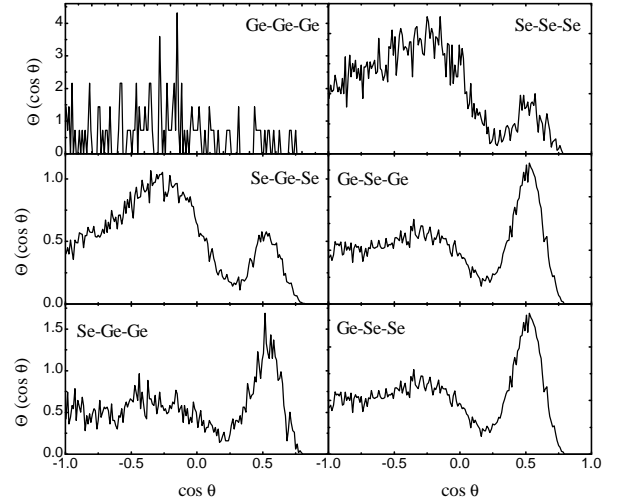}
\caption{\label{fig5} $\Theta_{ijl}(\cos\theta)$ functions obtained from RMC simulations.}
\end{figure}

By defining the partial bond-angle distribution functions $\Theta_{ijl}(\cos\theta)$  
where $j$ is the atom in the corner we calculated the angular distribution of the bonds between 
first neighbor atoms. The six $\Theta_{ijl}(\cos\theta)$ functions are shown in Fig.~\ref{fig5}. 
All these functions were calculated considering as $r_{\text{max}}$ the position of the first 
minimum after the peak of the first shell ($r_{\text{max}}\approx 3.0$~\AA).

The $\Theta_{\text{Ge-Ge-Ge}}(\cos\theta)$ function is very noisy because of the very small number
of Ge-Ge pairs in the first shell, but it shows a tendency for angles around 100$^\circ$. 
The $\Theta_{\text{Se-Se-Se}}(\cos\theta)$ function has peaks at 55--61$^\circ$ and 
a broad distribution from 99 to 118$^\circ$, with a maximum at 103$^\circ$. The internal Se-Se-Se 
angles in perfect tetrahedra are found at $60^\circ$, the Se-Se-Se angle in trigonal Se 
is seen at 103$^\circ$ and the angles in small Se chains and rings \cite{Kohara} can be found at 
90--116$^\circ$. Thus, the $\Theta_{\text{Se-Se-Se}}(\cos\theta)$ function indicates that the 
tetrahedra in MA-{\em a}-Ge$_{30}$Se$_{70}$ are slightly distorted and Se chains and rings 
are formed, in agreement with RS results and with the previous analyses of the $G_{ij}(r)$ functions.
The $\Theta_{\text{Se-Ge-Se}}(\cos\theta)$ function is very similar to the 
$\Theta_{\text{Se-Se-Se}}(\cos\theta)$ function, showing peaks at 58$^\circ$ and 105$^\circ$, 
which is close to the ideal tetrahedral angle of 109$^\circ$.

The $\Theta_{\text{Ge-Se-Ge}}(\cos\theta)$ function peaks at about 58$^\circ$ and 106$^\circ$. 
The Ge-Se-Ge sequence in ES units has angles around 80$^\circ$ that are not seen in this function, 
reinforcing the small quantity of these units. On the other hand, this sequence in 
CS units has angles around 
100$^\circ$, and $n$-fold rings have angles ranging from 92$^\circ$ to 125$^\circ$.\cite{Cobb2,Kohara}
The $\Theta_{\text{Se-Ge-Ge}}(\cos\theta)$ and $\Theta_{\text{Ge-Se-Se}}(\cos\theta)$ functions 
are similar to the others, having peaks at 58$^\circ$, associated with threefold rings, and 
at 116$^\circ$ (Se-Ge-Ge) and 109$^\circ$ (Ge-Se-Se), which are related to tetrahedral angles and 
$n$-fold rings. The $\Theta_{ijl}(\cos\theta)$ functions above confirm that distorted tetrahedral 
units are formed in MA-{\em a}-Ge$_{30}$Se$_{70}$, with a clear preference for CS 
units. These 
units seem to be connected by Se-Se bridges, forming small chains and rings, as pointed out by 
the RS data.

\section{Conclusion}
\label{secconclusion}

To summarize, we can conclude that the amorphous Ge$_{30}$Se$_{70}$ alloy can be produced 
by MA starting from the elemental powders of Ge and Se, but the structure of the alloy 
is different from that found in MQ-, VE- or MG-GeSe$_2$ samples, making clear the importance 
of the preparation technique. Structural units similar to distorted CS and 
ES tetrahedra are formed, with a preference for CS tetrahedra, as indicated by Raman spectroscopy. 
These units seem to be connected by Se-Se bridges, as suggested by the high number of these pairs 
in the first shell, by the increase in the distance between Ge-Ge second neighbors, 
by the $\Theta_{ijl}(\cos \theta)$ 
functions and also by RS vibrational data. These differences in the SRO affect the IRO of the alloy, 
and this causes the low intensity of the FSDP in ${\cal S}(K)$ of  
MA-{\em a}-Ge$_{30}$Se$_{70}$ when compared to the MQ-GeSe$_2$ alloy \cite{Petri,Salmon2}. The low 
intensity of the FSDP in ${\cal S}(K)$ can be traced back to the 
partial ${\cal S}_{\text{Ge-Ge}}(K)$ factor, whose FSDP is related to the ES units, which are 
found in a small quantity in the alloy. ${\cal S}_{\text{CC}}(K)$ reflects these 
features and shows a very weak FSDP when compared to the factor found for {\em l}-GeSe$_2$ 
\cite{Carlo,Carlo2,Penfold} or MQ-GeSe$_2$ \cite{Salmon2}. 

As a second remark, this study reinforces the relevance of using and combining RMC simulations 
with other techniques to model amorphous structures, since all features described above were 
obtained considering directly the experimental ${\cal S}(K)$ in the simulations.

\acknowledgments

We would like to thank the Brazilian agencies CNPq, CAPES and FAPESP for financial support. 
We wish to thank Dr. P\'al J\'ov\'ari (HASYLAB) for helpful suggestions about the RMC simulations. 
We are indebted to Dr. Philip Salmon (University of Bath) for sending the data on MQ-GeSe$_2$.

%\bibliographystyle{apsrev}
%\bibliography{ge30se70}

\end{document}